\documentclass[prd,twocolumn,amsmath]{revtex4}
\usepackage{amssymb,amsmath}
\usepackage{graphicx}
\usepackage{color}
\usepackage{xfrac}

\usepackage[caption=false]{subfig}

\usepackage[dvipsnames]{xcolor}

\begin{document}
\title{Comment on ``Hadamard states for a scalar field in anti-de Sitter spacetime
with arbitrary boundary conditions''}

\author{J. P. M. Pitelli}
\email[]{pitelli@ime.unicamp.br}
\affiliation{Departamento de Matem\'atica Aplicada, Universidade Estadual de Campinas, 13083-859, Campinas, S\~ao Paulo, Brazil}
\altaffiliation[Also at ]{The Enrico Fermi Institute, The University of Chicago, Chicago, IL}

\begin{abstract}

In a recent paper (Phys. Rev. D 94, 125016 (2016)), the authors argued that the singularities of  the two-point functions on the Poincar\'e domain of the $n$-dimensional anti-de Sitter spacetime ($\text{PAdS}_n$) have  the Hadamard form, regardless of  which (Robin) boundary  condition is chosen at the conformal boundary. However, the argument  used to prove this statement was based on an incorrect expression for the two-point function $G^{+}(x,x')$, which was obtained by demanding $\text{AdS}$ invariance for the vacuum state. In this comment I show that their argument works only for Dirichlet and Neumann boundary conditions and that the full $\text{AdS}$ symmetry cannot be respected by nontrivial Robin conditions (i.e., those which are neither Dirichlet nor Neumann). By studying the conformal scalar field on $\text{PAdS}_2$, I find the correct expression for $G^{+}(x,x')$ and show that, notwithstanding this problem, it still have the Hadamard form.
\end{abstract}

\maketitle

In a seminal paper~\cite{allen}, Allen and Jacobson presented a method of finding  two-point functions  in maximally symmetric spacetimes. Their method was based on the assumption that the state $|\psi\rangle$ is maximally symmetric. Within  this assumption,  the two point functions $G(x,x')$ constructed using $|\psi\rangle$ depends on $x$ and $x'$ only upon the geodesic distance $\sigma(x,x')$. The wave equation $(\square-m^2-\xi R)\varphi(x)=0$ then implies that $G(\sigma)$ satisfies the  differential equation (my notation agrees with that of Ref.~\cite{dappiaggi})
\begin{equation}
u(1-u)\frac{d^2G(\sigma)}{du^2}+\left[c-(a+b+1)u\right]\frac{dG(\sigma)}{du}-abG(\sigma)=0,
\label{eq z}
\end{equation}
where $u$ is related to the geodesic distance $\sigma$ by $u=\cosh^2{\left(\frac{\sqrt{2\sigma}}{2}\right)}$ (for anti-de Sitter spacetime) and 
\begin{equation}
\begin{aligned}
&a=\frac{n-1}{2}-\nu,\\
&b=\frac{n-1}{2}+\nu,\\
&c=n/2.
\end{aligned}
\label{abc}
\end{equation}
In Eq.~(\ref{abc}), $n$ is the spacetime dimension and $\nu=\frac{1}{2}\sqrt{1+4\tilde{m}^2}$, with $\tilde{m}^2\equiv m^2+\left(\xi-\frac{n-2}{4(n-1)}\right) R$, where $m$ represents the mass parameter and $\xi$  is the  scalar-curvature coupling constant. A convenient pair of linear independent solutions of Eq.~(\ref{eq z}) is given by
\begin{equation}
\begin{aligned}
&\left(1/u\right)^a {}_2F_1(a,a-c+1;a-b+1;1/u),\\
&\left(1/u\right)^b  {}_2F_1(b,b-c+1;b-a+1;1/u),
\end{aligned}
\label{two independent}
\end{equation}
with  ${}_2F_1$ being the Gauss' hypergeometric function. Clearly, any  linear combination of the solutions in Eq.~(\ref{two independent}) will be $\text{AdS}$ invariant. In Ref.~\cite{dappiaggi}, it was argued that the Green's function $G^{+}(x,x')=\langle 0|\varphi(x)\varphi(x')|0\rangle$, constructed from a field $\varphi$ satisfying a general Robin boundary condition, can be represented by such  a linear combination. My claim in this comment is that this assumption is  in general incorrect,  being true only  for Dirichlet and Neumann boundary conditions. In this way, except for these two particular boundary conditions, $G(x,x')$ will not be maximally symmetric. 

To illustrate my previous observations, let me focus on conformal fields on $\text{PAdS}_2$, since   a closed form for the two-point function can be easily derived in this case. The metric on $\text{PAdS}_2$ has the form
\begin{equation}
ds^2=\frac{-dt^2+dz^2}{z^2},\,\,\,z>0,
\end{equation}
with the conformal boundary located at $z=0$. The  geodesic distance satisfies the relation  $u=1+\frac{\sigma_M}{2z z'}$, where
\begin{equation}
\sigma_{M}=\frac{1}{2}\left[-(t-t')^2+(z-z')^2\right].
\end{equation}

For conformal fields in two dimensions we must have $\tilde{m}^2=0$, so that the general solution for $G(x,x')$ is  given by
\begin{equation}\begin{aligned}
G(x,x')&=A  \,{}_2F_1(0,0;0;1/u)\\&+B \left(1/u\right)\, {}_2F_1(1,1;2;1/u).
\end{aligned}
\label{general}
\end{equation}
At the conformal boundary we have $u\to\infty$. Therefore, in this limit we have
\begin{equation}\begin{aligned}
G(x,x')&\sim A  +B \left(1/u\right)\\&=A+ \frac{4Bz z'}{-(t-t')^2+(z+z')^2}.
\end{aligned}\end{equation}
Notice that if $A=0$, then $G(x,x')$ satisfies the Dirichlet boundary condition
\begin{equation}
G(t,z=0;t',z')=G(t,z;t',z'=0)=0,
\end{equation}
while if $B=0$, $G(x,x')$ satisfies the Neumann boundary condition
\begin{equation}
\frac{\partial G(t,z=0;t',z')}{\partial z}=\frac{\partial G(t,z;t',z'=0)}{\partial z'}=0.
\end{equation}
If we try to impose that $G(x,x')$ satisfy Robin boundary condition at the conformal boundary, i.e., that
\begin{equation}\begin{aligned}
&G(t,z=0;t'z)-\beta\frac{\partial G(t,z=0;t',z')}{\partial z}=0,\\
&G(t,z;t',z'=0)-\beta \frac{\partial G(t,z;t',z'=0)}{\partial z}=0.
\end{aligned}
\end{equation}
then $A$ and $B$ would satisfy
\begin{equation}
A-\frac{ B\beta z'}{-(t-t')^2+z'^2}=A-\frac{B\beta z}{-(t-t')^2+z^2}=0. 
\end{equation}
This does not make sense since they are constant. We hence conclude that for general Robin boundary condition ($\beta\neq 0$), the two-point function $G$ does not satisfy Eq.~(\ref{eq z}). Therefore the vacuum $|0\rangle$ cannot be maximally symmetric.  Notice that a length scale is introduced in the semiclassical theory when $\beta$ is finite and non zero. This extra length scale is responsible for the break of $\text{AdS}$ invariance. Clearly, this is not the case for Drichlet and Neumann boundary conditions.

In Ref.~\cite{dappiaggi}, it was correctly proved that $G(x,x')$ has the Hadamard form for Dirichlet and Neumann boundary conditions. The argument was then  extended to generic Robin boundary conditions by simply taking the linear combination of the fundamental solutions above. However, this is not correct as I showed above. 

The only thing left to do in the conformal case is to find the correct expression for $G(x,x')$ in the case  $\beta\neq 0$. In order to do so, I use the mode sum method, which is correct with or without additional symmetries. It can be  easily checked that the complete set of solutions $\{u_{\omega}^{(\beta)}(x)\}$ of the wave equation
\begin{equation}
\square\varphi(x)=-\frac{\partial^2\varphi(x)}{\partial t^2}+\frac{\partial^2\varphi(x)}{\partial z^2}=0,
\end{equation}
which satisfy Robin boundary conditions, and orthogonal in the Klein-Gordon inner product, is given by
\begin{equation}
u_{\omega}^{(\beta)}(t,z)=\frac{1}{\sqrt{\pi\omega}}\frac{\sin{\omega z}+\beta\omega \cos{\omega z}}{\sqrt{1+\beta^2\omega^2}} e^{-i\omega t}.
\end{equation}
As in Ref.~\cite{dappiaggi}, I choose to work with the Green's function $G^{+}(x,x')$. As a sum of modes, it is  given by
\begin{equation}\begin{aligned}
&G^{+}(x,x')=\langle 0|\phi(x)\phi(x')|0\rangle\\&=\int_0^\infty\frac{d\omega}{\pi\omega}\frac{(\sin{\omega z}+\beta\omega \cos{\omega z})(\sin{\omega z'}+\beta\omega \cos{\omega z')}}{1+\beta^2\omega^2} \\ &\times e^{-i \omega (t-t')-\epsilon\omega}.
\end{aligned}\label{integral}
\end{equation}
Notice that the  Robin boundary condition for $z$ and $z'$ is trivially satisfied in this case.

The integral~(\ref{integral}) can be exactly calculated, and is found to be
\begin{widetext}
\begin{equation}\begin{aligned}
&G^{(+)}(x,x')=\Bigg\{\frac{1}{2\pi} e^{\frac{-\Delta t+z+z'+i \epsilon }{\beta }} \text{E}_1\left(-\frac{(1-i \omega \beta ) (\Delta t-z-z'-i \epsilon )}{\beta }\right)+\frac{1}{2\pi}  e^{\frac{\Delta t+z+z'-i \epsilon }{\beta }} \text{E}_1\left(\frac{i (-i+\omega \beta ) (\Delta t+z+z'-i \epsilon )}{\beta }\right)\\&+\frac{1}{4\pi}\left[-\text{E}_1(-i \omega (-\Delta t+z+z'+i \epsilon ))+\text{E}_1(i \omega (\Delta t+z-z'-i \epsilon ))+\text{E}_1(i \omega (\Delta t-z+z'-i \epsilon ))-\text{E}_1(i \omega (\Delta t+z+z'-i \epsilon )\right]\Bigg\}_0^{\infty}.
\end{aligned}
\end{equation}
\end{widetext}
In  the above expression, $E_1(z)$ is the the exponential integral defined by
\begin{equation}
E_1(z)=\int_{z}^{\infty}{\frac{e^{-t}}{t}dt}.
\end{equation}
By using the asymptotic expansions for $\text{E}_1$ given by~\cite{abramowitz}
\begin{equation}\begin{aligned}
\text{E}_1(z)\sim-\gamma-\log{z}, \,\,|z|<<1\\
\text{E}_1(z)\sim \frac{e^{-z}}{z}, \,\, |z|\to\infty,
\end{aligned}
\end{equation}
 we arrive at
\begin{widetext}
\begin{equation}\begin{aligned}
&G^{(+)}(x,x')=\frac{1}{2\pi} e^{\frac{\Delta t+z+z'-i \epsilon }{\beta }} E_1\left(\frac{\Delta t+z+z'-i \epsilon }{\beta }\right)+\frac{1}{2\pi} e^{\frac{-\Delta t+z+z'+i \epsilon }{\beta }} E_1\left(-\frac{\Delta t-z-z'-i \epsilon }{\beta }\right)\\&+\frac{1}{4\pi}\left[\log (\Delta t-z-z'-i \epsilon)-\log (\Delta t+z-z'-i \epsilon)-\log (\Delta t-z+z'-i \epsilon)+\log( \Delta t+z+z'-i \epsilon)\right].
\end{aligned}
\label{dirichletmaisnao}
\end{equation}
\end{widetext}
Notice that the first two terms in Eq.~(\ref{dirichletmaisnao}) are regular in the limit $\Delta t\to 0$ and $z' \to z$. Moreover, the last term satisfies the wave equation and the Dirichlet boundary condition. Let us concentrate on the second term: a simple calculation shows that
\begin{equation}
G^{(+)}_{\text{Dirichlet}}=\frac{1}{4\pi}\log{\left(1-\frac{1}{\cosh^2{\left(\frac{\sqrt{2\sigma}}{2}\right)}}\right)},
\end{equation} 
so that in the limit $\sigma\to 0$ we have
\begin{equation}
G^{(+)}_{\text{Dirichlet}}\sim \frac{1}{2\pi}\left(\log{\sigma}-\log{2}\right),
\end{equation}
which has the  expected Hadamard form.

In summary:  although the $\text{AdS}$ spacetime is maximally symmetric,  the vacuum state does not respect its symmetries, except for fields satisfying Dirichlet or Neumann boundary conditions. In spite of that, the two-point function $G^{+}(x,x)$ thus has the expected Hadamard form for all Robin boundary conditions. In the above example, this happened because the Green's function could be separated into one term respecting Dirichlet boundary condition and one term depending on the boundary condition parameter $\beta$  with the last term being completely regular in the coincidence limit.  For more general situations - possibly non-conformal fields on $\text{PAdS}_n$ - we could, in principle, expand the mode sum in terms of powers of the boundary condition parameter $\beta$. The zeroth order contribution will satisfy Dirichlet boundary condition and  respect $\textrm{AdS}$ symmetries. Therefore, it will certainly have the required Hadamard form. We then expect that the remaining terms are regular when $\sigma\to 0$. This is subject of working in progress~\cite{pitelli}.

\acknowledgments

I am indebted to Professor R.~M. Wald, G.~Satishchandran, V.~S.~Barroso and Professor R.~A.~Mosna for clarifying several questions during the development of this comment and also thank the Enrico Fermi Institute for the kind hospitality. Finally, I thank Funda\c c\~ao de Amparo \`a Pesquisa do Estado de S\~ao Paulo (FAPESP) (Grants No. 2018/01558-9 and 2013/09357-9) for financial support.

\end{document}